# Thermodynamics and energy conversion of near-field thermal radiation: maximum work and efficiency bounds

Ivan Latella, Agustín Pérez-Madrid and J. Miguel Rubi[a]

Departament de Física Fonamental, Facultat de Física, Universitat de Barcelona, Martí i Franquès 1, 08028 Barcelona, Spain

**Abstract.** We analyse the process of conversion of near-field thermal radiation into usable work by considering the radiation emitted between two planar sources supporting surface phonon-polaritons. The maximum work flux that can be extracted from the radiation is obtained taking into account that the spectral flux of modes is mainly dominated by these surface modes. The thermodynamic efficiencies are discussed and an upper bound for the first law efficiency is obtained for this process.

## 1 Introduction

Thermal radiation produced by perfect emitters opened the door to new physics at the beginning of the past century, and the properties of blackbody radiation are reviewed in many classic textbooks from the point of view of both thermodynamics [1] and statistical mechanics [2]. In the past years, with the arrival of new technologies, the control of devices working at the nanoscale has been possible and efforts are currently made to understand and predict the physics associated to this scale. For characteristic scales smaller than the thermal wavelength of the radiation emitted by the components of such devices, interference effects become relevant and the optical properties of the involved materials play an important role [3, 4]. Due to optical properties of the material, the radiative heat transfer in these scales can notably exceed the blackbody limit [5-10]. For small separations between two bodies, evanescent waves that rapidly decay near the interface of the materials become the main contribution in the radiative heat transfer, leading to what can be interpreted as a tunnelling of photons between the bodies [4]. As a consequence, the number of contributing modes of electromagnetic radiation for a given frequency is drastically modified with respect to the corresponding one for bodies which lie further apart from each other. The information about the contribution of different modes is codified in the density of states and the spectral flux of modes, which are the physical quantities one uses to obtain thermodynamic functions of the radiation [11-13].

The enhancement of the radiative heat transfer in the near-field offers also new possibilities regarding conversion of thermal energy into work using thermophotovoltaic cells [14, 15]. Recently, both the possibility of implementing advanced materials such as graphene as emitters in near-field thermophotovoltaic devices [16], and a hybrid graphene-semiconductor device which consists of a photovoltaic cell covered by a graphene sheet [17], have been anticipated.

Our aim here is to discuss the thermodynamics of the thermal radiation energy-conversion process by considering the radiation between two planar sources in the near-field regime. In particular, we consider the thermal sources are polar materials supporting surface phonon-polaritons, whose radiation in the near-field regime is strongly dominated by these surface waves.

---

[a] Corresponding author: mrubi@ub.edu

## 2 Maximum work and thermodynamic efficiency

In an energy-conversion process, the converter transforms an input energy into usable work, and this input energy can be regarded as the initial internal energy of the system from which the work is being extracted. As a consequence of the energy conversion, the system under consideration ends in a new state with a different internal energy. In order to define the thermodynamic scheme of the process, let the internal energy per unit volume of the system be denoted by $u(T)$, $T$ being the temperature, and its variation by $\Delta u = u(T_e) - u(T_h)$, where $T_h$ and $T_e$ are the temperatures of the initial and of the final state, respectively. Let us also assume that in this final state the system is in thermal equilibrium with the environment (at temperature $T_e$), with $T_h > T_e$, and that the evolution of the system proceeds at constant volume. During the process of conversion, the energy satisfies the balance equation $\Delta u + q_e + w = 0$, where $q_e$ is the heat delivered to the environment and $w$ is the work extracted from the system. Here both $q_e$ and $w$, as well as the entropies below, are also taken to be per unit volume of the system. Furthermore, the formulation of the second law of thermodynamics [18] states that the entropy change in the system $\Delta s$ and the entropy change in the environment $\Delta s_e$ are related in such a way that $\Delta s + \Delta s_e \geq 0$. The variation of entropy in the system is given by $\Delta s = s(T_e) - s(T_h)$, where $s(T)$ is the entropy of the system at temperature $T$. For irreversible, stationary processes that take place in the time interval $\Delta t$, the average total entropy production per unit volume is defined by [19]

$$\left(\frac{ds_{irr}}{dt}\right)\Delta t \equiv \Delta s + \Delta s_e . \tag{1}$$

Thus, taking equation (1) and the energy balance equation into account, and using that $\Delta s_e = q_e/T_e$, one has

$$w = T_e \Delta s - \Delta u - T_e \left(\frac{ds_{irr}}{dt}\right)\Delta t . \tag{2}$$

Let us now introduce the ideal work $w_{ideal} = T_e \Delta s - \Delta u$, which is the work obtained in a process with no entropy production. Since the last term in the right-hand side of equation (2) is always negative or zero, the maximum work available from such a transformation is precisely given by the ideal work $w_{ideal}$.

Two different points of view can be adopted in the approach to the computation of the thermodynamic efficiency. On the one hand, one can consider the first law efficiency $\eta_I$, which is the ratio of the available work to the input energy. The input energy here is $u(T_h)$, which is the initial available energy. Hence,

$$\eta_I = \frac{w}{u(T_h)} = \frac{w_{ideal} - T_e(ds_{irr}/dt)\Delta t}{u(T_h)} . \tag{3}$$

One sees that an upper bound $\bar{\eta}_I$ for this efficiency is obtained by considering that the work is ideal. Therefore,

$$\bar{\eta}_I = \frac{w_{ideal}}{u(T_h)}, \tag{4}$$

so that $\bar{\eta}_I \geq \eta_I$. On the other hand, the second law efficiency $\eta_{II}$ measures how the system deviates from ideal operating conditions. It is given through [19]

$$\eta_{II} = \frac{w}{w_{ideal}} = 1 - \frac{T_e(ds_{irr}/dt)\Delta t}{w_{ideal}} . \tag{5}$$

Note that $\eta_{II} = \eta_I/\bar{\eta}_I$, as pointed out by Landsberg and Tonge [20], and we always have $\eta_{II} \geq \eta_I$. The most favourable situation is when there is no entropy production, so that the system works in ideal conditions and hence $\eta_{II} = 1$. Thus, no practical information can be obtained from $\eta_{II}$ in ideal conversion processes.

For a continuous transformation involving radiative phenomena, one has to consider fluxes of energy and entropy taken per unit time and surface. In addition, we assume the system to be in a stationary state in which the net energy and entropy fluxes in the converter vanish. For isotropic propagative radiation emitted over an hemisphere [2], the energy and entropy fluxes are given by $\dot{U}(T) = cu(T)/4$ and $\dot{S}(T) = cs(T)/4$, respectively, were $c$ is the speed of light in vacuum. These expressions for the fluxes involving the factor $c/4$ are only valid for propagative modes; below we will consider the case where evanescent waves also contribute. According to this, one also considers a work flux and a flux of heat delivered to the environment given by $\dot{W} = cw/4$ and $\dot{Q}_e = cq_e/4$, respectively. Moreover, the variations of energy and entropy fluxes of the radiation are given by $\Delta\dot{U} = \dot{U}(T_e) - \dot{U}(T_h)$ and $\Delta\dot{S} = \dot{S}(T_e) - \dot{S}(T_h)$. Accordingly, one has $\Delta\dot{U} + \dot{Q}_e + \dot{W} = 0$ and

$$\Delta\dot{S} + \Delta\dot{S}_e = \Delta\dot{S}_{irr} \geq 0, \tag{6}$$

where $\Delta\dot{S}_e = \dot{Q}_e/T_e$ is the variation of entropy flux of the environment and $\Delta\dot{S}_{irr} = \frac{c}{4}(ds_{irr}/dt)\Delta t$ is the entropy production flux due to irreversibilities in the process of conversion. In addition, from (2) one has

$$\dot{W} = T_e\Delta\dot{S} - \Delta\dot{U} - T_e\Delta\dot{S}_{irr} = \dot{\mathcal{W}} - T_e\Delta\dot{S}_{irr}, \tag{7}$$

where in the last step we have introduced the ideal work flux

$$\dot{\mathcal{W}} \equiv T_e\Delta\dot{S} - \Delta\dot{U}, \tag{8}$$

which is the maximum work flux that can be extracted from the radiation. Therefore, for the fluxes associated to the radiation, the upper bound for the first law efficiency reads

$$\bar{\eta}_I = \frac{\dot{\mathcal{W}}}{\dot{U}(T_h)}. \tag{9}$$

Since equations (6), (7), (8) and (9) are based on a balance of energy and entropy fluxes, they are also valid for non-propagative modes if the fluxes are properly defined.

## 3 Thermal radiation in the near-field regime

In order to implement the concepts of maximum work and efficiency for thermal radiation, we first make some general comments about the relevant thermodynamic functions for this case. The internal energy density (per unit volume) of thermal radiation at temperature $T$ can be written as

$$u(T) = \int_0^\infty d\omega \; \hbar\omega n(\omega, T)\rho(\omega), \tag{10}$$

where $n(\omega, T) = \left(e^{\hbar\omega/(k_B T)} - 1\right)^{-1}$ is the mean occupation number of photons in a mode of frequency $\omega$, $\rho(\omega)$ is the density of states with frequency $\omega$, and $\hbar$ and $k_B$ are the reduced Planck constant and Boltzmann's constant, respectively. The function $\rho(\omega)$ depends on the microscopic details of how the radiation is emitted by the body. The specific heat at constant volume can be obtained from (10) yielding

$$c_V = \frac{\partial u(T)}{\partial T} = \int_0^\infty d\omega \; k_B \left[\frac{\hbar\omega}{k_B T} n(\omega, T)\right]^2 e^{\frac{\hbar\omega}{k_B T}} \rho(\omega). \tag{11}$$

In deriving equation (11) we have assumed that the density of states does not depend on temperature. Thus, according to usual thermodynamic relations, the entropy density of the radiation is readily given by $s(T) = \int_0^T dT' c_V(T')/T'$. Making the change $x = \hbar\omega/(k_B T')$ in the integration over the temperature in the previous expression, taking into account (11), and introducing

$$m(\omega, T) = \int_{\hbar\omega/(k_B T)}^\infty \frac{x \, dx}{4\sinh^2(x/2)} = [1 + n(\omega, T)] \ln[1 + n(\omega, T)] - n(\omega, T) \ln n(\omega, T), \tag{12}$$

the entropy density takes the form

$$s(T) = \int_0^\infty d\omega \; k_B m(\omega, T) \rho(\omega). \tag{13}$$

Once the internal energy and entropy densities are determined, other thermodynamic potentials can be obtained via Legendre transformations.

The energy flux can also be defined by

$$\dot{U}(T) \equiv \int_0^\infty d\omega \; \hbar\omega n(\omega, T) \varphi(\omega), \tag{14}$$

which leads to an associated entropy flux that can be written as

$$\dot{S}(T) \equiv \int_0^\infty d\omega \; k_B m(\omega, T) \varphi(\omega). \tag{15}$$

The function $\varphi(\omega)$ is called the spectral flux of modes. If only propagative modes are considered, the spectral flux of modes is related to the density of states via $\varphi_{\text{prop}}(\omega) = c\rho(\omega)/4$. The case of blackbody radiation, which corresponds to propagative electromagnetic waves, is obtained by considering $\varphi(\omega) = c\rho_{\text{bb}}(\omega)/4$, with the blackbody density of states $\rho_{\text{bb}}(\omega) = \omega^2/(\pi^2 c^3)$. Taking this density of states into account, the energy and entropy fluxes become $\dot{U}_{\text{bb}} = \sigma T^4$ and $\dot{S}_{\text{bb}} = 4\sigma T^4/3$, respectively, where $\sigma$ is Stefan's constant. We stress that the above formulation does not depend on the specific theory underlying the radiative process, whose details are reduced to $\varphi(\omega)$. In order to incorporate the physical mechanism involved in the emission of quanta, the corresponding spectral flux of modes must be computed.

Now consider a body that emits thermal radiation on the surface of a second body that is separated by a vacuum gap of width $d$, and assume that both bodies have planar surfaces. Consider also that the temperature of the first body is $T_h$ and that the second body is in thermal equilibrium with the environment at temperature $T_e$, with $T_h > T_e$. The two bodies can be seen as sources of thermal radiation. We are interested in implementing a converter that transforms the energy flux of the radiation of the hot source into a usable work flux by considering that the cold source also emits thermal radiation. The situation is the same as discussed in the previous section with the system under consideration being the radiation that evolves from a state at temperature $T_h$ into a state at temperature $T_e$. Therefore, using (8) with (14) and (15), the ideal work flux reads

$$\dot{W} = \int_0^\infty d\omega \; \{k_B T_e [m(\omega, T_e) - m(\omega, T_h)] - \hbar\omega[n(\omega, T_e) - n(\omega, T_h)]\} \varphi(\omega). \tag{16}$$

Using the density of states of blackbody radiation $\rho_{\text{bb}}(\omega)$, which does not depend on the gap width $d$, this quantity takes the form

$$\dot{\mathcal{W}}_{\text{bb}} = \sigma(T_{\text{h}}^4 - T_{\text{e}}^4) - \frac{4}{3}\sigma(T_{\text{h}}^3 - T_{\text{e}}^3), \tag{17}$$

that, e.g., for $T_{\text{h}} = 330$ K and $T_{\text{e}} = 300$ K gives $\dot{\mathcal{W}}_{\text{bb}} \sim 10$ W m$^{-2}$. We will see that for near-field radiation emitted by real materials, the amount of work flux can be considerably increased.

The optical properties of the materials are introduced by taking into account the Fresnel reflection coefficients of each material-vacuum interface [3, 4], namely, $R_{i,\alpha}(\kappa, \omega)$, where $\kappa$ is the component of the wave vector parallel to the surfaces, the subscript $i = 1, 2$ refers to the bodies, and $\alpha = $ p, s refers to the two different polarizations. The reflection coefficients depend on the dielectric constants of the materials and hence contain the information about the electromagnetic fields leading to the emission of radiation. Thus, if the electromagnetic fields are known, using the fluctuation-dissipation theorem [21] and computing the average of the normal component of the Poynting vector, the radiated energy flux can be obtained [3, 4, 22, 23] and, therefore, the spectral flux of modes identified. According to this procedure, in this case the spectral flux of modes is given by

$$\varphi(\omega) = \frac{c}{4} \sum_{\alpha=\text{p,s}} \left\{ \int_0^{\omega/c} \frac{d\kappa\, \kappa}{c\pi^2} \frac{\left[1 - |R_{1,\alpha}(\kappa,\omega)|^2\right]\left[1 - |R_{2,\alpha}(\kappa,\omega)|^2\right]}{\left|1 - e^{2i\gamma d}R_{1,\alpha}(\kappa,\omega)R_{2,\alpha}(\kappa,\omega)\right|^2} \right. \\ \left. + \int_{\omega/c}^{\infty} \frac{d\kappa\, \kappa}{c\pi^2} \frac{4e^{-2|\gamma|d}\text{Im}[R_{1,\alpha}(\kappa,\omega)]\text{Im}[R_{2,\alpha}(\kappa,\omega)]}{\left|1 - e^{-2|\gamma|d}R_{1,\alpha}(\kappa,\omega)R_{2,\alpha}(\kappa,\omega)\right|^2} \right\}, \tag{18}$$

where $\gamma = \sqrt{(\omega/c)^2 - \kappa^2}$. The first term in curly brackets in equation (18) corresponds to propagative waves, while the second term corresponds to the contribution of evanescent waves. The blackbody density of states is obtained from (18) by assuming that the bodies are perfect absorbers so that $R_{i,\alpha}(\kappa, \omega) = 0$. Hence, in this case only propagative modes contribute to the spectral flux of modes. In contrast, when the optical properties of the surfaces are taken into account and the separation $d$ between them is much smaller than the thermal wavelength $\lambda_T = \hbar c/(k_B T)$, that is, 7.6 μm for $T = 300$ K, the spectral flux of modes is strongly dominated by the contribution of evanescent modes [3, 4]. This is the so-called near-field regime.

Here we consider polar materials that support surface phonon-polaritons, which are p-polarized modes of the interface that appear if the dielectric constant $\varepsilon(\omega)$ satisfies $\text{Re}[\varepsilon(\omega_0)] = -1$ for some frequency $\omega_0$ which is the frequency of this surface mode [23]. In the near-field regime, the radiation emitted by materials supporting surface phonon-polariton is highly monochromatic and the dominant frequency is $\omega_0$ [23, 24]. In addition, we consider two bodies of the same material and accordingly introduce $R_p(\omega) \equiv R_{i,p}(\omega)$, where we also have taken into account that in the electrostatic limit the reflection coefficient does not depend on $\kappa$. Thus, under these conditions and using the near-monochromatic approximation, the spectral flux of modes can be written as [13]

$$\varphi_{\text{nf}}(\omega) = g_d(\omega)\delta(\omega - \omega_0), \quad g_d(\omega) \equiv \frac{\text{Re}\left[\text{Li}_2\left(R_p^2(\omega)\right)\right]}{4\pi d^2 f'(\omega)}, \tag{19}$$

where $f(\omega) = \text{Im}[R_p^2(\omega)]/\text{Im}^2[R_p(\omega)]$, $\text{Li}_2(z)$ is the dilogarithm function, and the prime denotes the derivative with respect to $\omega$. To derive the form of the spectral flux of modes (19), the method used in [24] to compute the heat transfer coefficient has been employed. Using the near-field spectral flux of modes $\varphi_{\text{nf}}(\omega)$, the thermodynamic functions of interest can be obtained in this regime. From (14) and (15), the energy and entropy fluxes in the near field take the form

$$\dot{U}_{\text{nf}}(d, T) = \hbar\omega_0 n(\omega_0, T)g_d(\omega_0), \tag{20}$$

and

$$\dot{S}_{\rm nf}(d,T) = k_{\rm B} m(\omega_0, T) g_d(\omega_0), \tag{21}$$

respectively. Note that the fluxes behave as $1/d^2$ in this regime, as can be seen from equation (19). As a result, the ideal work flux (16) can be written as [13]

$$\dot{W}_{\rm nf}(d, T_{\rm h}, T_{\rm e}) = \hbar\omega_0 g_d(\omega_0) \left\{ \frac{k_{\rm B} T_{\rm e}}{\hbar\omega_0} [m(\omega_0, T_{\rm e}) - m(\omega_0, T_{\rm h})] - [n(\omega_0, T_{\rm e}) - n(\omega_0, T_{\rm h})] \right\}. \tag{22}$$

In order to illustrate the magnitude of $\dot{W}_{\rm nf}$ with an explicit example, we consider that the material of the sources is hexagonal boron nitride (hBN). The dielectric constant of this material can be described by the Lorentz model

$$\varepsilon(\omega) = \varepsilon_\infty \left( 1 + \frac{\omega_{\rm L}^2 - \omega_{\rm T}^2}{\omega_{\rm T}^2 - \omega^2 - i\Gamma\omega} \right), \tag{23}$$

where $\varepsilon_\infty$, $\omega_{\rm L}$, $\omega_{\rm T}$, and $\Gamma$ are material-dependent parameters that we take from [17]. For hBN the frequency of the surface phonon-polariton is $\omega_0 = 2.96 \times 10^{14}$ s$^{-1}$, so that taking $T_{\rm h} = 330$ K and $T_{\rm e} = 300$ K, and setting the separation between surfaces to $d = 20.6$ nm, one obtains a work flux $\dot{W}_{\rm nf} \sim 862$ W m$^{-2}$, almost two orders of magnitude more than blackbody radiation (see above). To achieve the same amount of work flux with blackbody radiation, the temperature of the hot source would need to be raised to $T_{\rm h} = 500$ K. In order to make a comparison of fluxes in the near-field with fluxes in blackbody regime, we define the reduced energy and entropy fluxes $\Upsilon(d, T) \equiv \dot{U}_{\rm nf}(d, T)/\dot{U}_{\rm bb}(T)$ and $\Sigma(d, T) \equiv \dot{S}_{\rm nf}(d, T)/\dot{S}_{\rm bb}(T)$, respectively, and the reduced maximum work flux $\Phi(d, T_{\rm h}, T_{\rm e}) \equiv \dot{W}_{\rm nf}(d, T_{\rm h}, T_{\rm e})/\dot{W}_{\rm bb}(T_{\rm h}, T_{\rm e})$. In Figure 1 these reduced fluxes are shown as a function of $d$ for sources made of hBN.

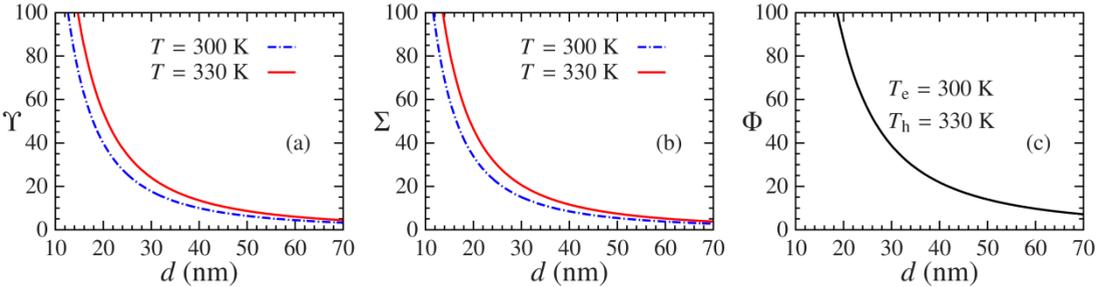

**Figure 1.** Reduced (dimensionless) fluxes as a function of the gap width $d$ considering surfaces made of hBN. (a) Reduced energy flux for two different temperatures. (b) Reduced entropy flux for two different temperatures. (c) Reduced maximum work flux taking 300 K as environmental temperature and 330 K for the temperature of the hot source.

For blackbody radiation, the bound for the efficiency (9) is given by [20]

$$\bar{\eta}_{\rm bb} = 1 - \frac{4}{3}\frac{T_{\rm e}}{T_{\rm h}} + \frac{1}{3}\left(\frac{T_{\rm e}}{T_{\rm h}}\right)^4. \tag{24}$$

In contrast, since near-field radiation is strongly dominated by the frequency of the resonant mode, one expects that the upper bound for the efficiency (9) corresponds to that of near-monochromatic radiation [20]. This is indeed the case, and (9) becomes [13]

$$\bar{\eta}_{\rm nf} = 1 - \frac{n(\omega_0, T_{\rm e})}{n(\omega_0, T_{\rm h})} + \frac{k_{\rm B} T_{\rm e}}{\hbar\omega_0} \frac{m(\omega_0, T_{\rm e}) - m(\omega_0, T_{\rm h})}{n(\omega_0, T_{\rm h})}. \tag{25}$$

It can be shown that for fixed temperatures $\bar{\eta}_{nf}$ increases for increasing $\omega_0$ and approaches the Carnot efficiency $\eta_C = 1 - T_e/T_h$ as $\omega_0$ goes to infinity. In addition, depending on the frequency of the surface phonon-polariton, $\bar{\eta}_{nf}$ can be higher than $\bar{\eta}_{bb}$ for a certain range of temperatures, as shown in Figure 2 for hBN. Moreover, when the difference of temperatures of the sources is small, this bound for the efficiency increases as the temperature of the environment is decreased.

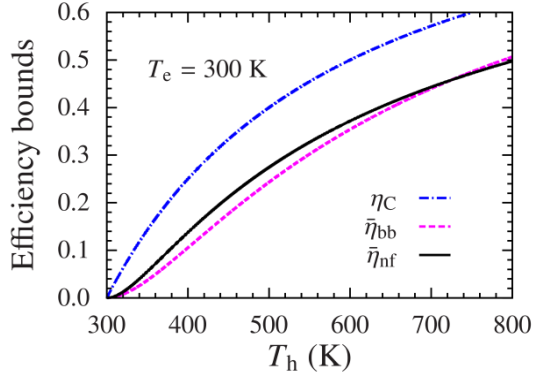

**Figure 2.** Efficiency bounds as a function of the temperature of the hot source $T_h$ for fixed $T_e = 300$ K. The near-field efficiency $\bar{\eta}_{nf}$ is plotted for hBN, and for comparison also the Carnot efficiency $\eta_C$ is plotted.

## 4 Discussion

We have analysed the process of conversion of thermal radiation energy into usable work using a thermostatistical approach and discussed upper bounds for the first law efficiency in different regimes. Special attention has been paid to thermal radiation in the near-field regime considering that the sources are polar materials supporting surface phonon-polaritons. In particular, we have presented a comparison between energy, entropy, and ideal work fluxes for hBN and the corresponding fluxes for blackbody radiation. We have seen that for a relative small difference of temperatures, with respect to typical room temperature, the maximum work flux that can be obtained from near-field radiation is almost two orders of magnitude more than the corresponding for blackbody radiation if one considers sources of hBN. For other materials with a lower resonance frequency, e.g., silicon carbide, the maximum work flux is even higher. The present treatment can be applied to a wide range of materials that support these surface waves, which includes many semiconductors.

## Acknowledgements

This work was supported by the Spanish Government, under Grant No. FIS2011-22603. I.L. acknowledges financial support through an FPI Scholarship (Grant No. BES-2012-054782) from the Spanish Government and J.M.R. acknowledges financial support from Generalitat de Catalunya under program ICREA Academia.